# A Simulation-based Education Approach for the Electromagnetic and Electromechanical Transient Waves in Power Systems

Abdulelah Alharbi[1], Shutang You[1], Yilu Liu[1,2]
1. University of Tennessee, Knoxville; 2. Oak Ridge National Laboratory
abdulelah009@gmail.com; syou3@utk.edu; liu@utk.edu

*Abstract*— **Power systems usually go through electromagnetic and electromechanical transient processes after different disturbances. Learning the characteristics and the differences between them are important but not easy for students majoring in power systems. This paper presents a simulation-based approach to comprehensively study the two types of transient waves, constituting the experimental part of the power system transient and stability course. In this approach, three models with different levels of complexity are developed to simultaneously show the two types of transient waves in the time domain. The developed models are then demonstrated as testbeds for investigating various aspects related to the two types of transient, such as waveforms induced by different disturbances, influencing factors on the propagation speed, and the interaction between incident waves and the reflective waves. In addition, a theory-to-practice engineering research process is demonstrated through developing a power system event-location application, which is inspired by electromechanical wave propagation study. The proposed education process and models at various complexity levels provide a creative and interactive way for power system transients and dynamics study.**

## I. Introduction

Power grid reliability is important since it ensures continuous power supply for customers. Many power grid simulations at the system levels are based on electromechanical simulations, for example, frequency response studies. The wide-area monitoring of power grids also mainly focuses on electromechanical dynamics. As the increase of renewable generation, power system modeling and simulation of electromagnetic dynamics is becoming increasingly important for understanding the dynamics of power grids.

In fact, power grid disturbances in power systems would generate electromagnetic and electromechanical transient waves that propagate over the whole system from the disturbance location [1, 2]. The two transient waves are not isolated processes and they have mutual influences from the beginning of the disturbances. However, due to their differences in duration, model complexity, and study focuses, most research models and studies the two transient waves separately.

Electromagnetic transients describe the process of energy exchange between the electric filed and the magnetic field, as well as the corresponding changes in current and voltage. Electromagnetic waves describe the propagation of electromagnetic transients from the disturbance location to the whole system. The study time scale of electromagnetic transients is from milliseconds to seconds after the disturbance, while the propagation velocity of an electromagnetic transient wave is at the same magnitude as the speed of light. In electromagnetic transient simulations, the system circuit model is needed to obtain the time-domain system response. Equations in simulation include component characteristics equations, which could be differential equations, and topology constraints such as KCL and KVL. Line models have a significant impact on electromagnetic simulation. Most recent electromagnetic transient wave programs apply the travelling wave models of transmission lines [3], such as the Bergeron Model and Frequency Dependent Models, which is seldom used in electromechanical transient simulation. Electromagnetic transient simulation can help to assess the impact of lighting and switching surge, protection device selection and deployment, fault location, and mitigate electromagnetic interference caused by overvoltage in power systems.

Unlike electromagnetic transients, electromechanical transients also incorporate energy exchange between the kinetic energy of rotors and the energy in electric/magnetic fields. One feature of electromechanical transients is that there is significant change in rotor speed. The time scale of electromechanical transient study is from one to dozens of seconds, while the propagation speed of electromechanical waves is around 200-1000 miles per second, much lower than that of electromagnetic waves. For relatively simple chain systems, some research has derived the wave equations and calculated the theoretical speed of electromechanical wave propagation. [4] regarded generators and transmission lines as a continuous system and derived standard second-order wave equations to describe the electromechanical wave [5]. A series of applications have been developed based on electromechanical waves, such as fault location [6], oscillation detection analysis [7, 8], generation-load mismatch detection [9], etc.

Distinguishing and understanding the characteristics of the two transient waves is important in understanding the dynamic behaviors of power systems. Since both transient waves involve relatively complex theories, conventional course teaching methods introduce and investigate the two transient waves independently. Moreover, most of the developed cases focus on only one of the two transients. However, the abstract theories and independent simulation cases usually make students confused about the two transient waves, making it difficult to understand the dynamic behaviors of power systems overall.

This paper proposed a simulation-based approach to study these two transient waves. This approach includes creating a simulation model and disturbance that can show the electromagnetic wave and the electromechanical wave simultaneously in the time domain. Directly observing and comparing the two waves after one disturbance event in the time domain allows students to get a comprehensive and vivid impression. Based on this simulation model, some additional investigation study can be easily implemented through studying the wave changes after tuning the model or disturbance, such as the system inertia, transmission line impedance, and disturbance types. Other advantages of this simulation-based study approach include: 1) Simulating the two waves in one system helps to distinguish them easily and shows their impact on each other. 2) The propagation speed of the system can be calculated theoretically and also obtained in the simulation results. This allows a comparison and validation between the theory and simulations. 3) Sensitivity analysis of the two waves on system parameters and disturbances can help to investigate the impact factors of the system dynamic responses. 4) This approach allows students to compare real transient measurements with simulation observations. 5) This simulation-based approach can serve as a platform for testing new ideas and developing new application prototypes for system security. It can also serve as materials to teach students general simulation-based innovation methods.

## II. Electromechanical and Electromagnetic Waves

Several different events in large power system such as generation trips and load trips can cause a mismatch between the input mechanical power and the output electrical power of the generators which leads the generators' rotors to be accelerated with respect to their synchronous reference frame [3]. This movement of rotors can be described as electromechanical wave propagation. A sudden phase shift in one bus causes the flow of active power on the lines connected to the bus to be changed. Thus the angle difference between this bus and its adjacent buses changes as well which results in changing active power flow to other buses according to [10]:

$$P_{ij} = \frac{V_i V_j}{X_{ij}} \sin(\Delta\delta_{ij}) \tag{1}$$

Generally, a sudden phase disturbance on a certain bus affects the flow of active power in the entire power system and the disturbance is propagated from one bus to another in this way; so, the disturbance travels across the grid. This could lead to losing synchronization of several generators and result in negatively impact the reliable power transfer limits of the system [11].

One of the goals to study the electromechanical wave propagation is to estimate how fast is the wave travels and when it arrives at each location. It is important to note that the electromechanical wave propagation speed is affected by the machines mechanical parameters mainly generator inertia constant and transmission line inductive reactance. The expression for the propagation speed of the electromechanical wave as derived in [3]:

$$v = \sqrt{\frac{\omega V^2 \sin\theta}{2h|z|}} \tag{2}$$

where ω is the nominal system angular frequency, V is the amplitude of source voltage in per unit, θ is the phase angle of transmission line impedance in radian, h is the inertia constant per unit length in seconds, and z is per unit line impedance per unit length. When using this equation, the system with zero generator impedance gives better simulation results. For high voltage transmission line, the line impedance phase angle θ is approximately $\frac{\pi}{2}$ [3].

Electromagnetic transient happens due to a sudden change in the energy stored in capacitors in the form of electric field and in inductors in the form of magnetic field. Some random events such as lightning, switching events, and network resonances and harmonics from power electronic devices cause changes in voltages and currents. Capacitors and inductors can support the voltage or current effectively by keeping the pre-disturbance states as the stored energy either electrical or mechanical is discharged. This causes a delay in disturbance propagation as it moves along the line [12]. Therefore, the electromagnetic speed can be calculated using only the transmission line parameters; especially, inductance and capacitance of the line. The velocity of propagation for a lossless line is [13]:

$$v = \frac{1}{\sqrt{LC}} \tag{3}$$

When the internal flux linkage of the conductor is neglected, $GMR_L = GMR_C$

$$v = \frac{1}{\sqrt{\mu_0 \varepsilon_0}} \cong 3 \times 10^8 \, m/s \tag{4}$$

Typical extra high voltage lines will have propagation speeds from $2.5 \times 10^8 m/s$ to $2.91 \times 10^8 m/s$ [12].

## III. Simulation Models

### A. *The Ring Bus System*

In order to calculate the speed of both electromagnetic and electromechanical waves, a 23-bus ring system was constructed as shown in Fig.1. Each bus of the system has one generator and one load as can be seen in Fig.2. Each generator is rated at power of 120 MW while the load with rating of 1000 MW. The system is made up of a ring of transmission lines with impedance of 0.325 ohm/km. The internal voltage magnitudes of all generators are assumed to be 1 per unit and zero phase angle over the entire network.

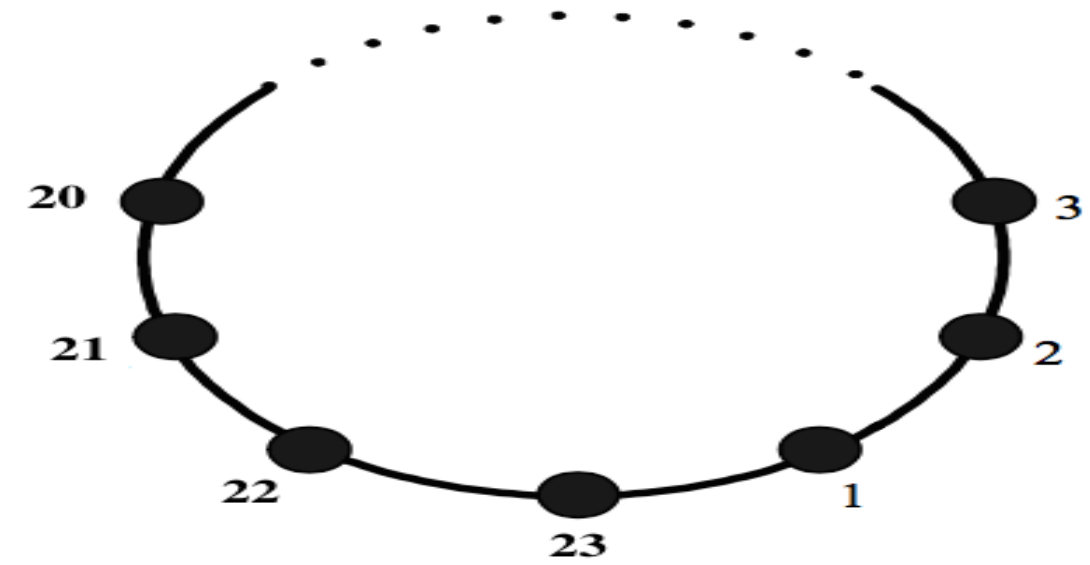

Figure 1: The 23-Bus Ring System

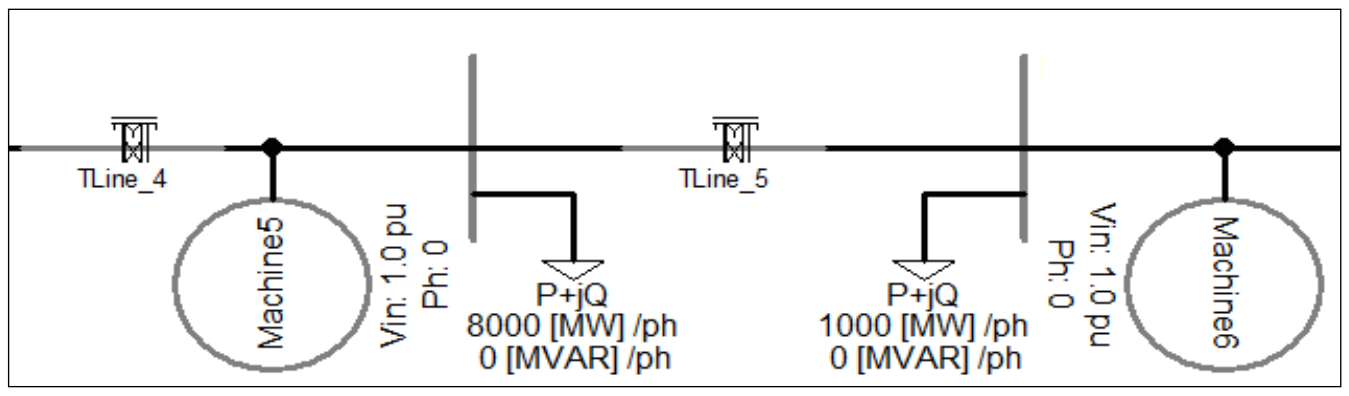

Figure 2: The Configuration of Each Bus in the 23-Bus System

### *B. The 23-Bus System Model*

The 23-bus system, shown in Figure 3, has six generators with total generation of 3400 MVA and seven connected loads which rating of 3538 MVA. The system also has four shunt capacitors with rating of 906 Mvar and one shunt reactor with rating of 614 Mvar. Additionally, the system has ten transformers. All necessary data of the system including bus data, generator data, line data, load data, shunt data, and transformer data are provided in PSS/E examples [14].

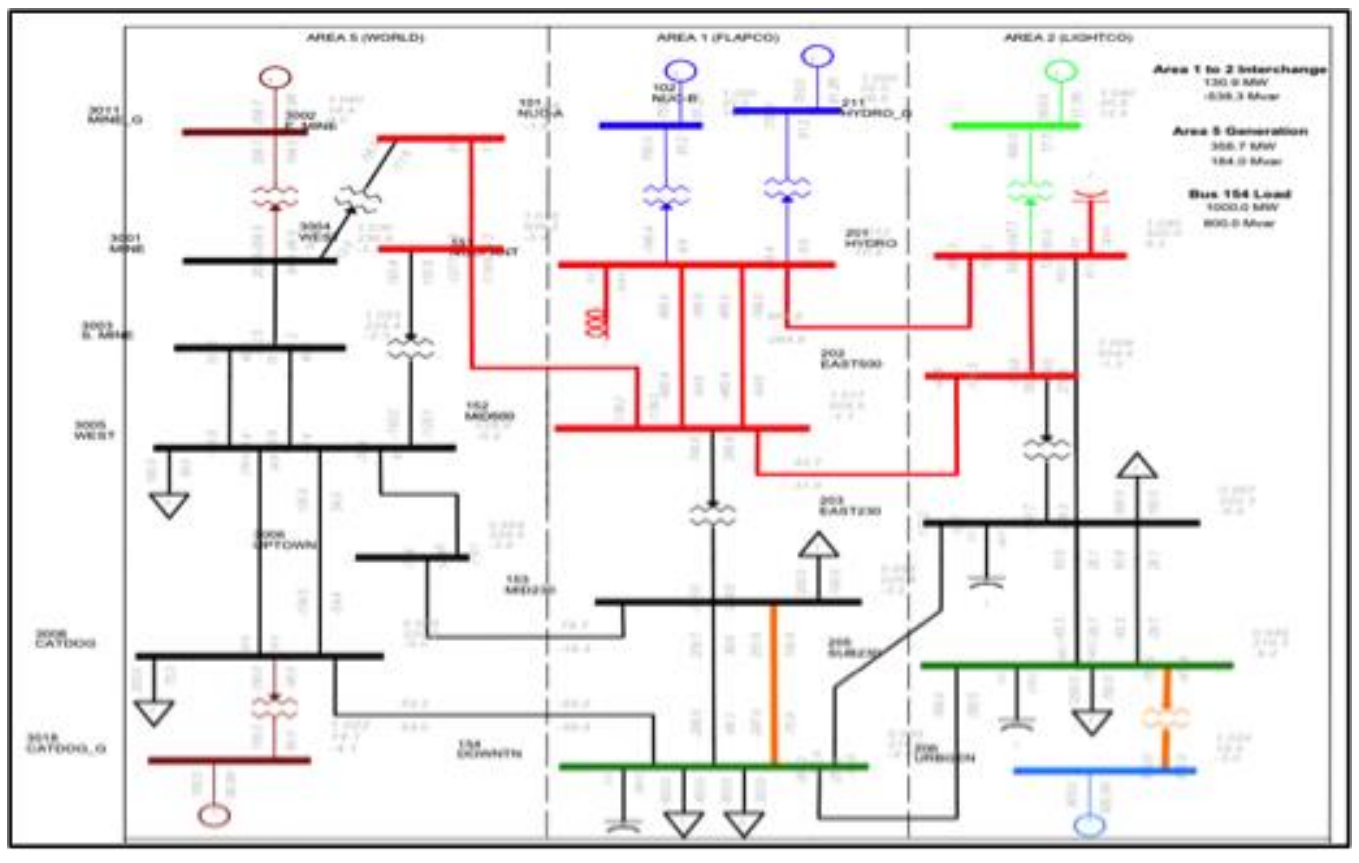

Figure 3: The 23-Bus System Model

### *C. The Simplified NPCC System Model*

The ring bus system is an ideal system for theoretical study. In practice, most systems are networks. The second simulation model is constructed based on the NPCC system in the Eastern Interconnection of U.S. With its interfaces presented by generators, this model has 48 generators and 206 branches. Fig.4 shows the NPCC system diagram.

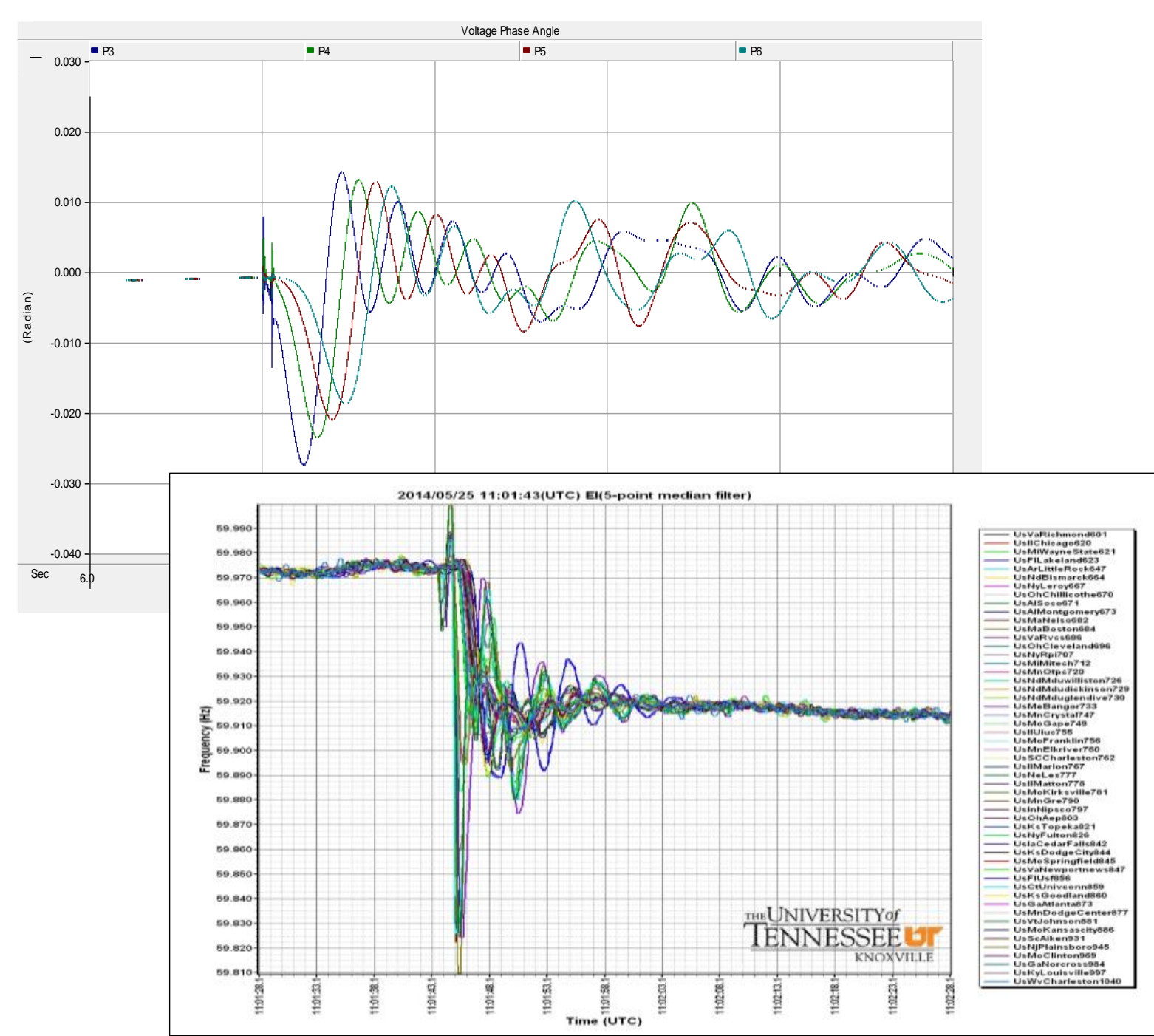

Figure 5 (a) Electromagnetic and Electromechanical Transient

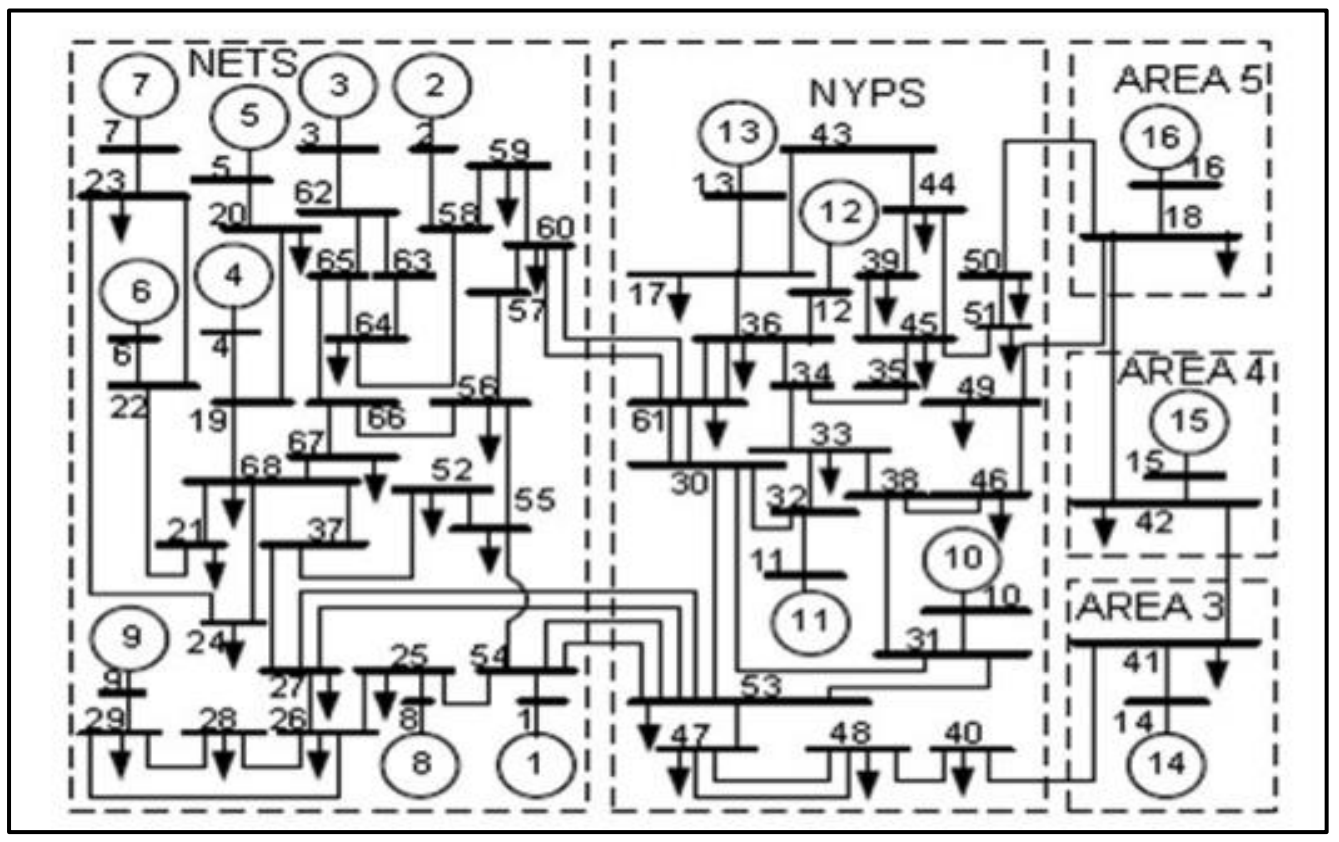

Figure 4: The NPCC System Model

## IV. Recommended Simulations and Observations

### A. *Distinguish Electromagnetic and Electromechanical Transients*

A single line to ground fault is applied at Bus 1 for the 23-bus system. Fig.5 (a) shows both the electromagnetic and electromechanical transients at different buses. Fig.5 (b) zooms the electromagnetic transient in. The differences of the two transient are clearly shown in the time domain: electromagnetic transients form and decay quickly while electromechanical transient develop slow and have periodical features. Both transients have in common that the transients occur earlier and have a higher magnitude if the measurement point is closer to the disturbance.

Figure 5: (b) Electromagnetic Transient (zoomed in)

The Frequency Monitoring Network (FNET), which was proposed in 2001 by Virginia Tech and was established in 2004, is an Internet based, wide-area frequency monitoring system that measures the voltage phase angle, voltage amplitude, and frequency from a single-phase voltage source at distribution system level. GPS-time synchronized frequency disturbance recorder (FDR) is the main component of FNET system that can provide frequency information at different locations in a power grid. More information about FNET can be found in [6, 15-17]. It can provide valuable data about electromechanical transients. Fig. 6 displays electromechanical transient that is resulted from a typical generation tripping event.

Figure 6: Generation Trip Event

### B. *Propagation speed*

#### 1) *Theoretical calculation and simulation result*

The velocity of the electromechanical wave propagation can be calculated using equation (2), where the inertia constant (H): 10, rated power of each generator (G): 120MW, number of coherent generators (Coh): 100, number of generator groups (N): 23, transmission line parameters: (R= 0.325ohm/km, L= 0.102µH, and C= 0.115nF), length of each line: 100km. 100 MVA and 500kV are used as base values.

$$h = \frac{H \times Coh \times N \times G}{MVAbase \times length \times (N-1)} = 12.55\ s/km$$

$$v = \sqrt{\frac{\omega V^2 \sin\theta}{2h|z|}} = 339.97\ km/s$$

For 500 kV transmission line, the electromagnetic wave propagation speed:

$$v = \frac{1}{\sqrt{LC}} \approx 2.93 \times 10^8 m/s$$

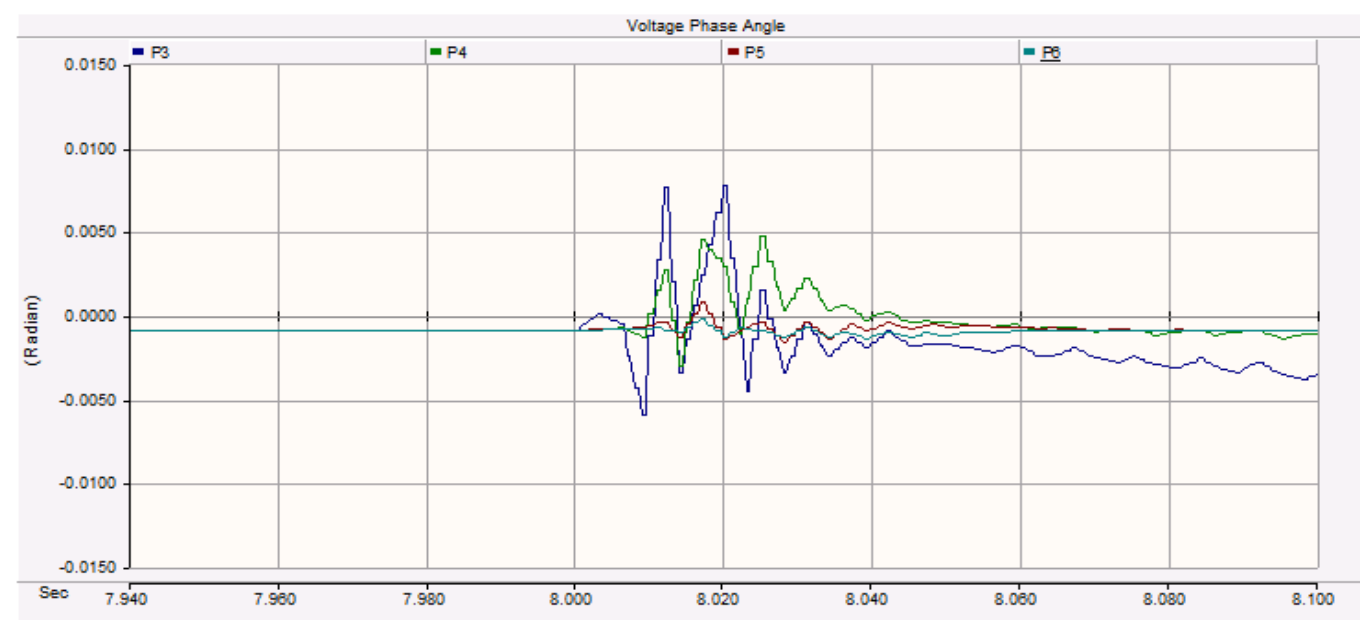

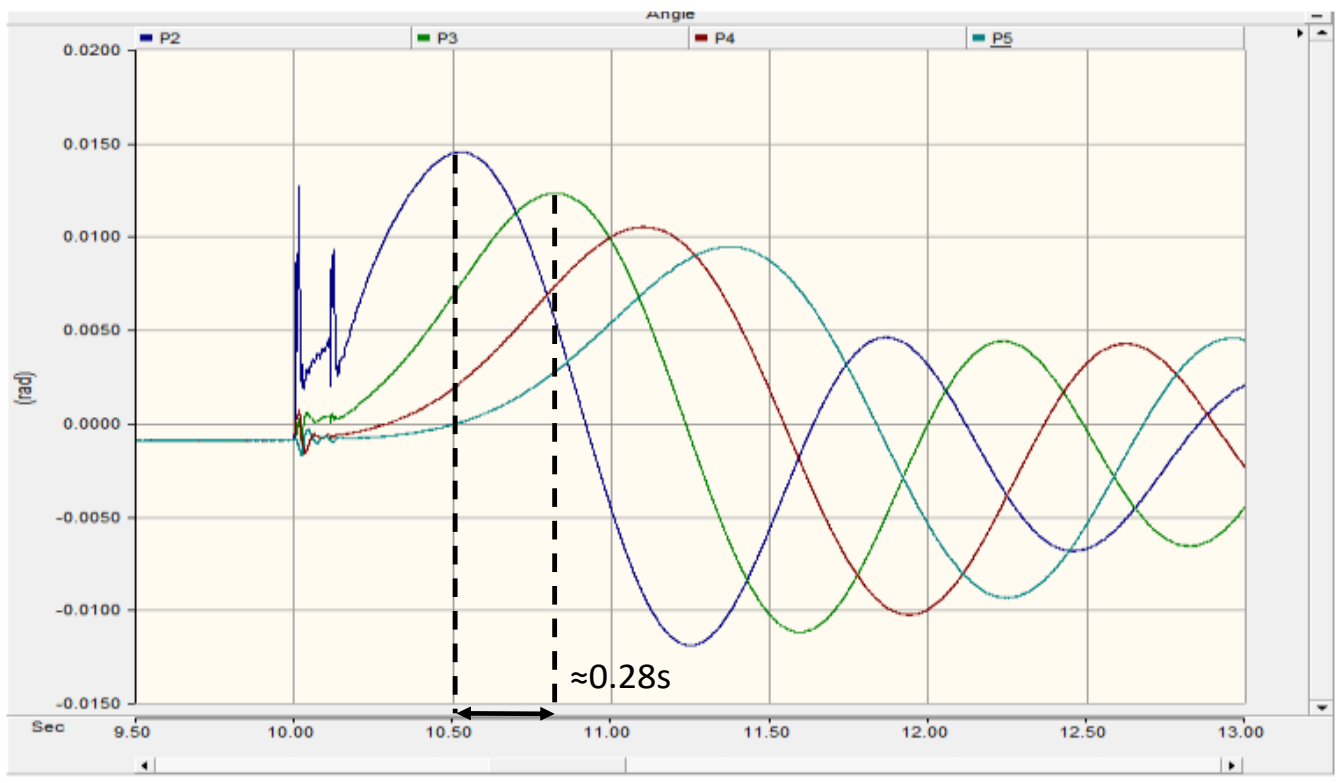

Figure 7: Electromechanical Wave Propagation Speed

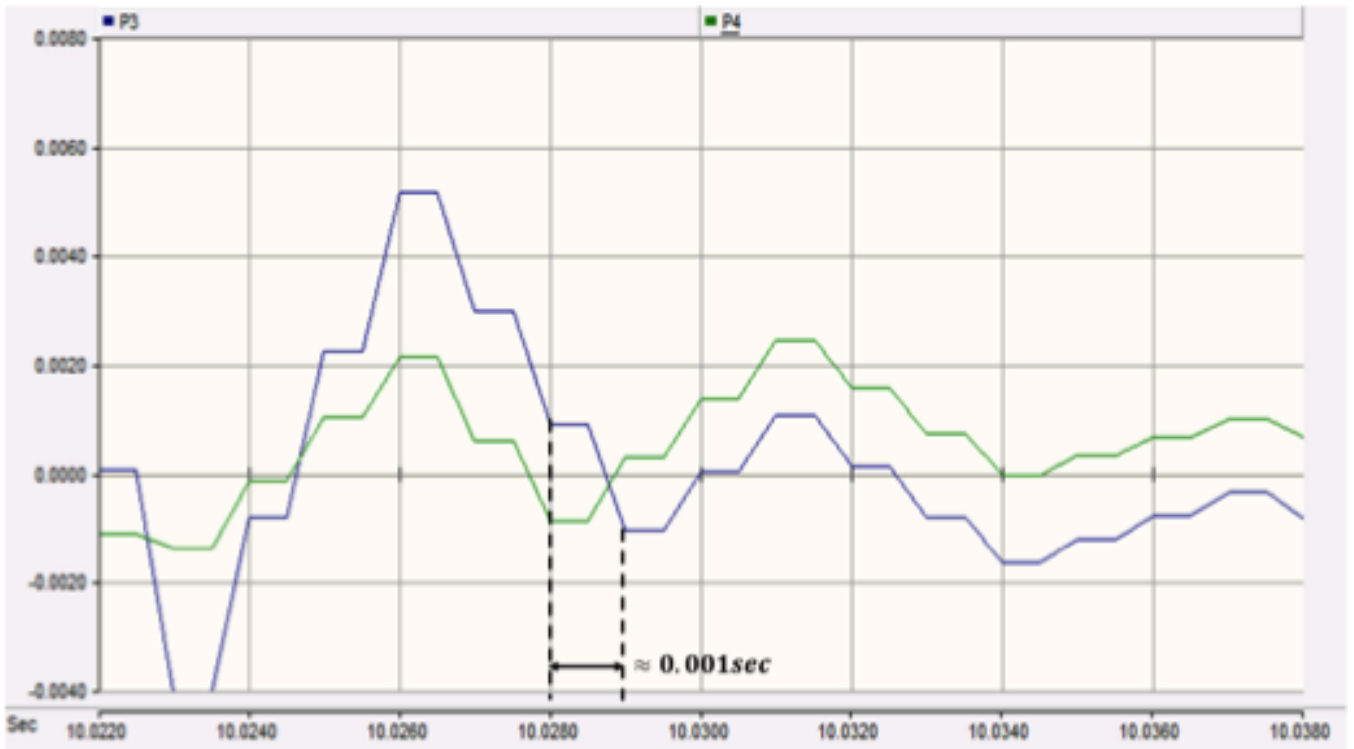

Figure 8: Electromagnetic Wave Propagation Speed

From the simulation, the speed of electromechanical wave propagation is estimated as $v = 100/0.28 = 357\ km/s$ as shown in Fig. 7 while the speed of electromagnetic wave propagation is estimated as $v = 200/0.001 = 2 \times 10^8\ m/s$ as shown in Fig. 8.

#### 2) *Electromechanical Waves: Generator Inertia Constant*

To investigate the impact of inertia constants of generators on electromechanical wave propagation speed, the inertia constants of the all generators are decreased from 10s to 5s. The speed of electromechanical wave propagation becomes 100km/0.2s=500 km/s as shown in Fig. 9. This shows that the electromechanical wave propagation speed increases as the inertia of the system decreases.

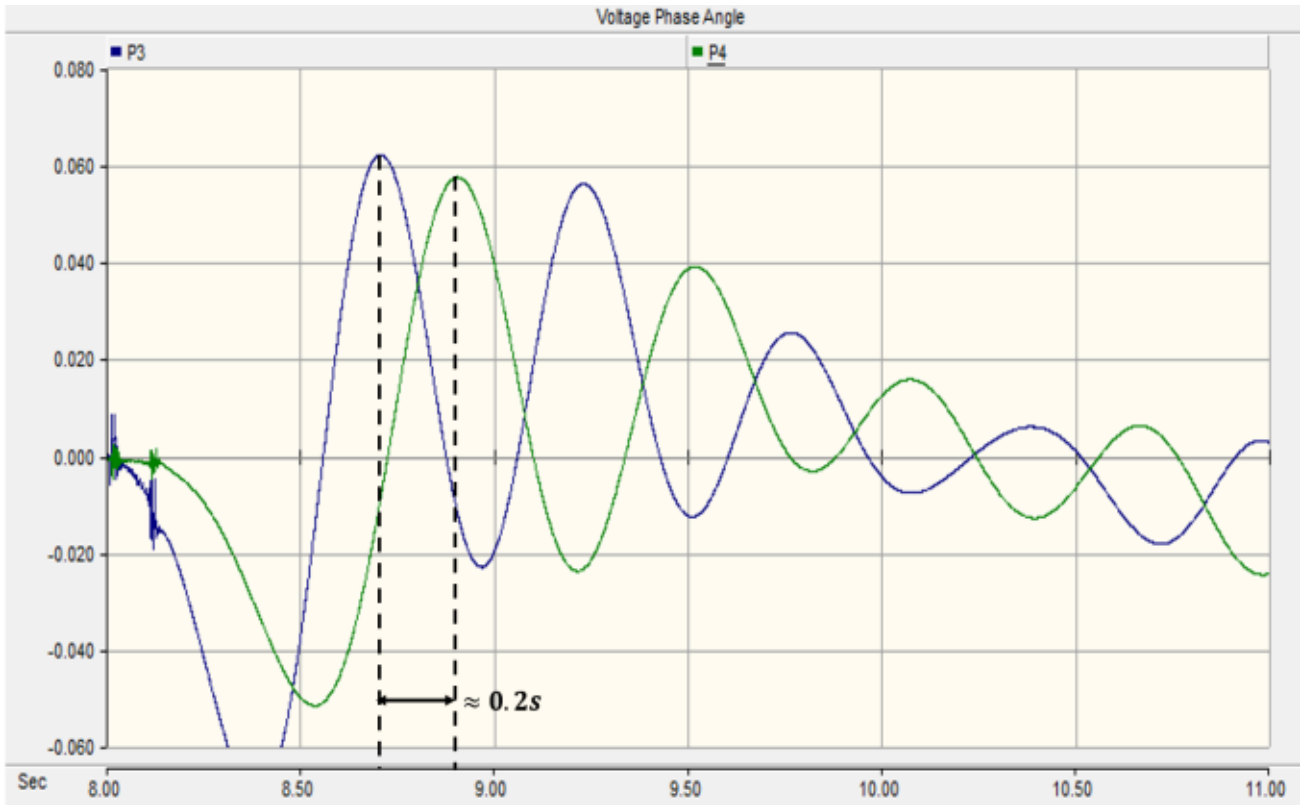

Figure 9: Wave Propagation Speed when the Inertia Constant is 5sec

### 3) *Electromechanical Waves: Transmission Line Reactance*

In order to study the influence of transmission line inductive reactance on the propagation speed of electromechanical waves, the inductance parameter of the transmission line increases from 0.102μH to 0.2μH. It can be noticed that the propagation speed of electromechanical waves decreases as the inductive reactance of the transmission lines increase. The speed of electromechanical wave propagation is $324.78\ km/s$ as shown in Fig 10.

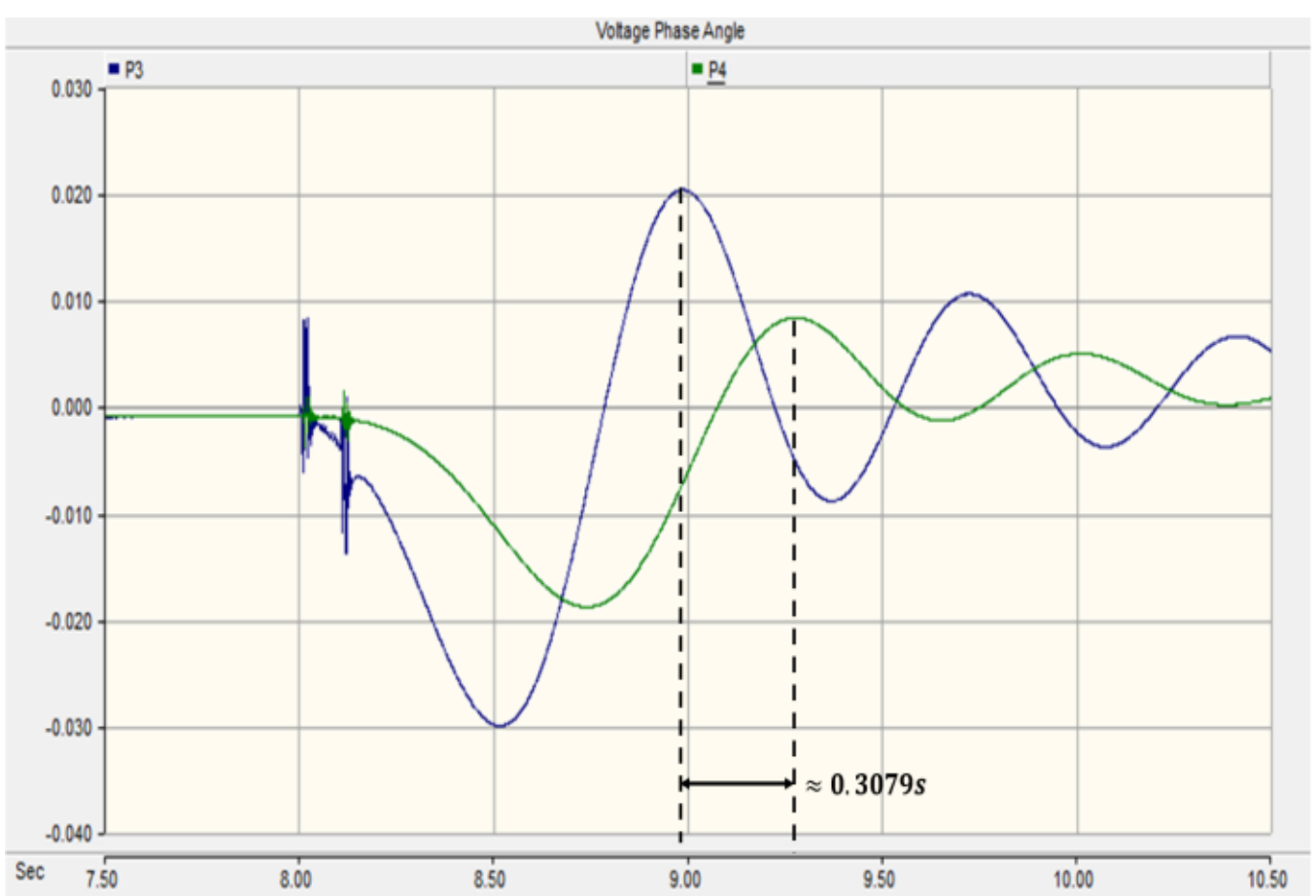

Figure 10: Wave Propagation Speed when the Line Inductance is 0.2μH

### 4) *Electromagnetic Waves: Transmission Line Capacitance*

Since the electromagnetic wave speed is associated with the transmission line parameters, increasing the transmission line capacitance would decrease the propagation speed as stated in the equation 3. Fig. 11 shows that the electromagnetic wave decreases to $1.11 \times 10^8\ m/s$ when the capacitance is increased to 0.8nF. The influence of increasing the capacitance value on the electromagnetic propagation speed is illustrated in Fig. 12. It can be clearly seen that the electromagnetic wave speed becomes slow as the capacitance values get high.

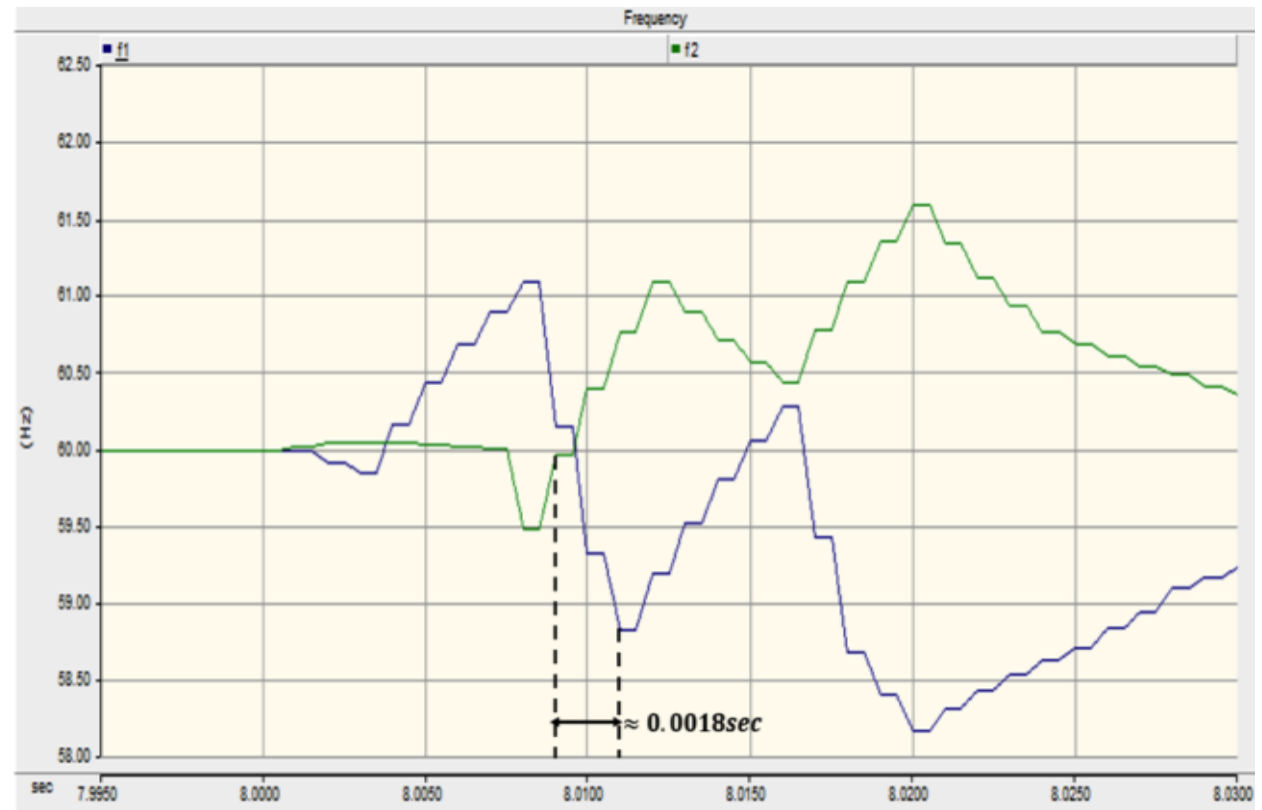

Figure 11: Wave Propagation Speed when the Line Capacitance is 0.8nF

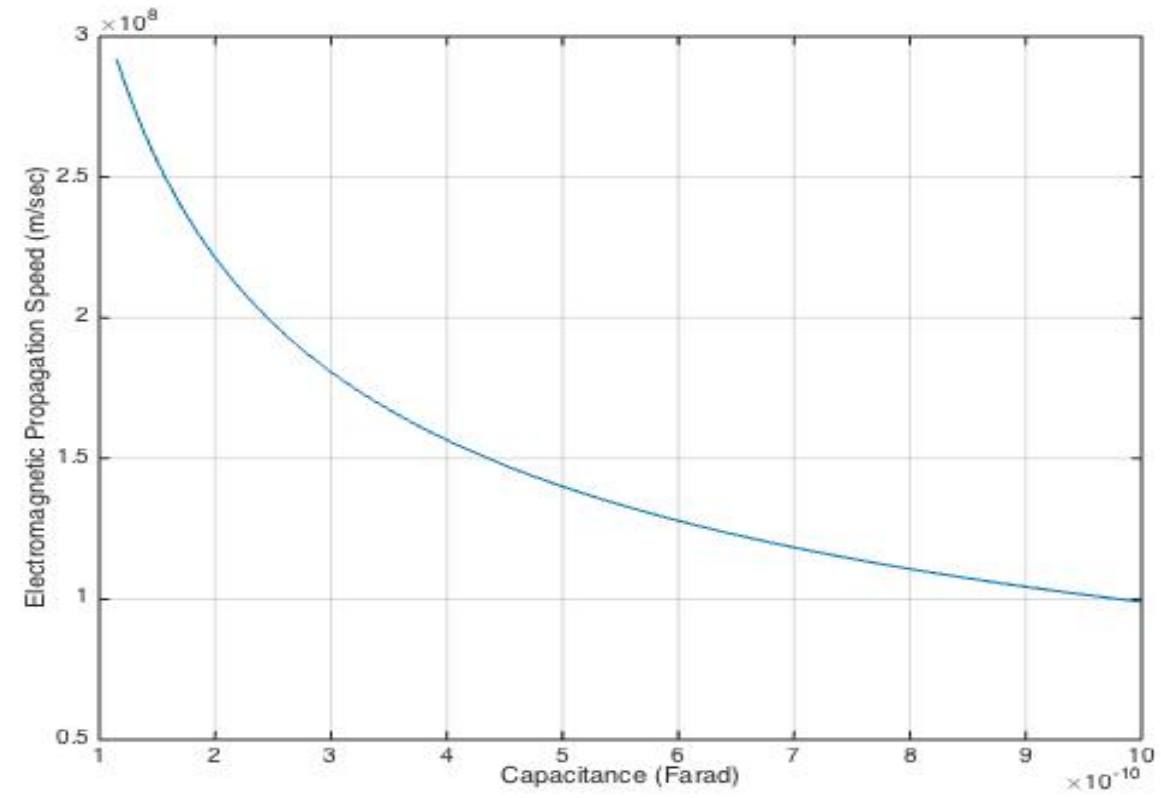

Figure 12: Relationship between Electromagnetic Speed and Line Capacitance

### 5) *Reflective Electromechanical Waves*

It is clear that for a ring system, waves travel in both directions from a faulted bus. They both meet at a specific instant of time as they travel at the same propagation speed. Thus, the electromechanical waves experience different distortion. It can be seen in Fig. 13 that the 23-bus closed loop system has a reflective electromechanical wave, which can be observed at around the 12.5s.

## C. *Event Types and Transients Waves*

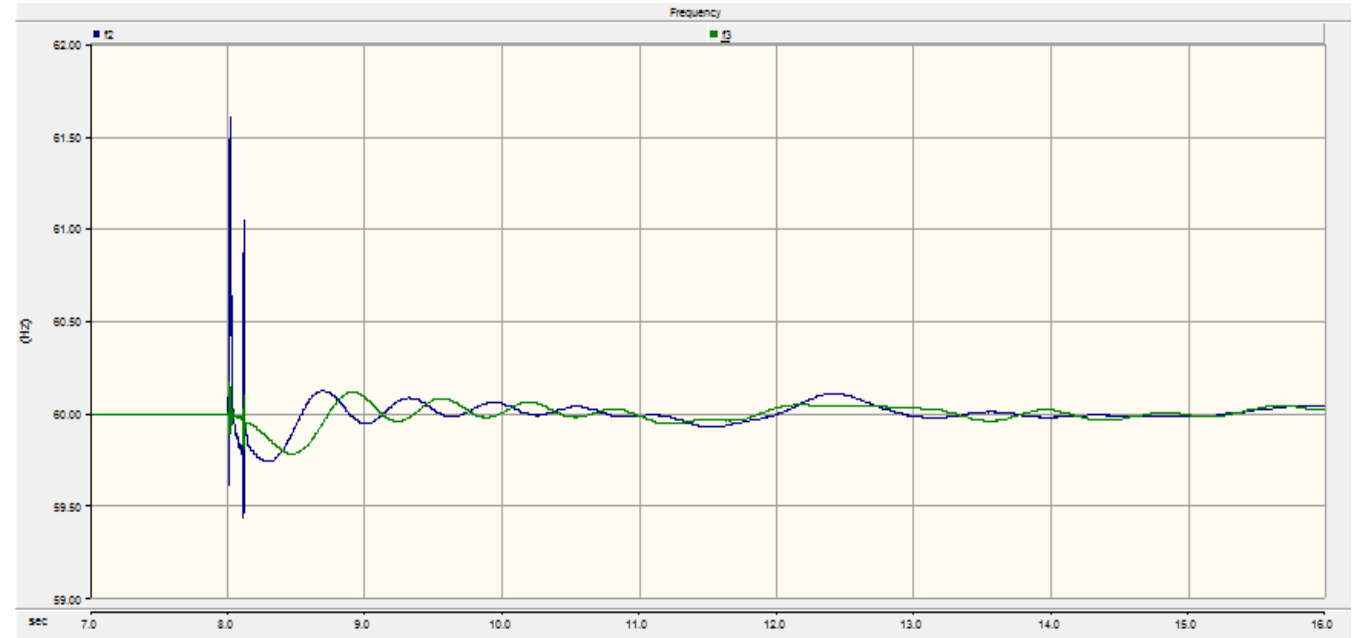

Figure 13: The Reflective Electromechanical Waves

Power system may be subjected to various types of disturbances, such as generation trip, line trip, load shedding, and faults. Different disturbances will generate electromagnetic and electromechanical waves with different characteristics. The electromechanical waves can help to identify the disturbance types and improve situational awareness, as well as to aid control and remedy decisions in power systems. Fig. 14 shows the electromagnetic and the electromechanical waves after different types of disturbances. It shows that the generation trip events cause a decrease wave of rotor speed, while load shedding events lead to an increase wave of the rotor speed.

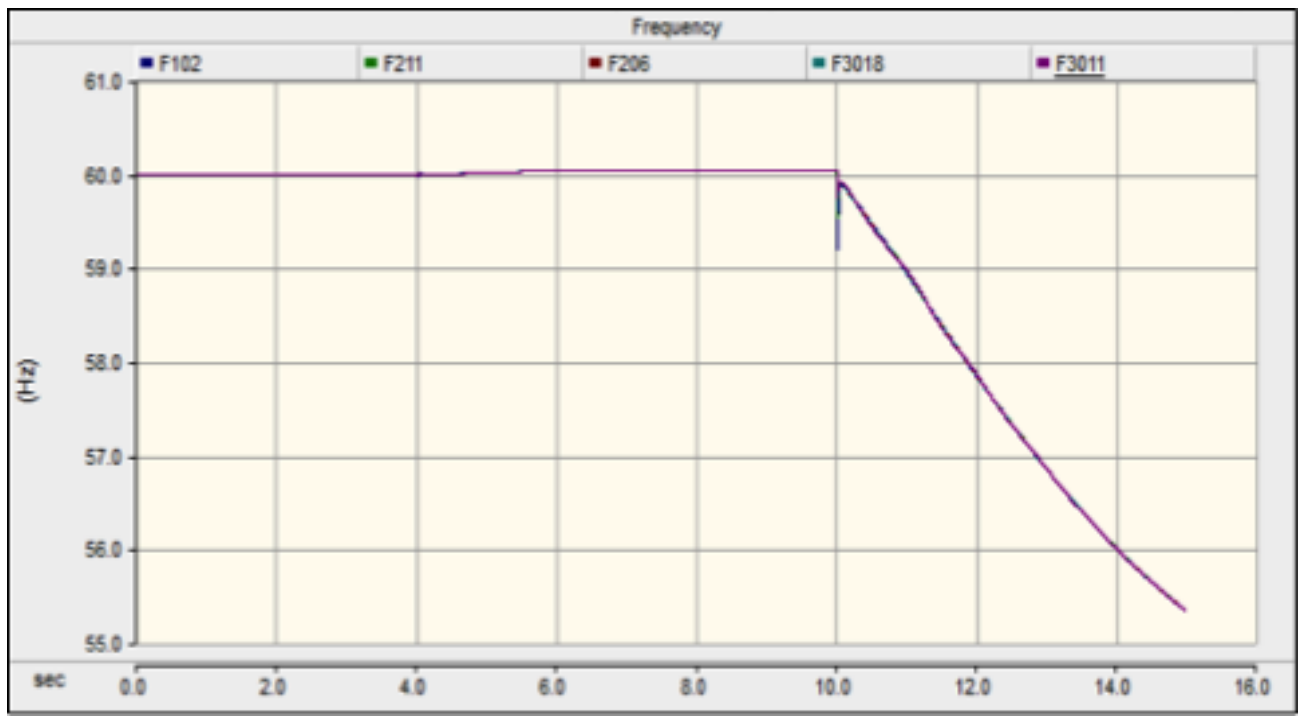

(A) Generation Trip Event

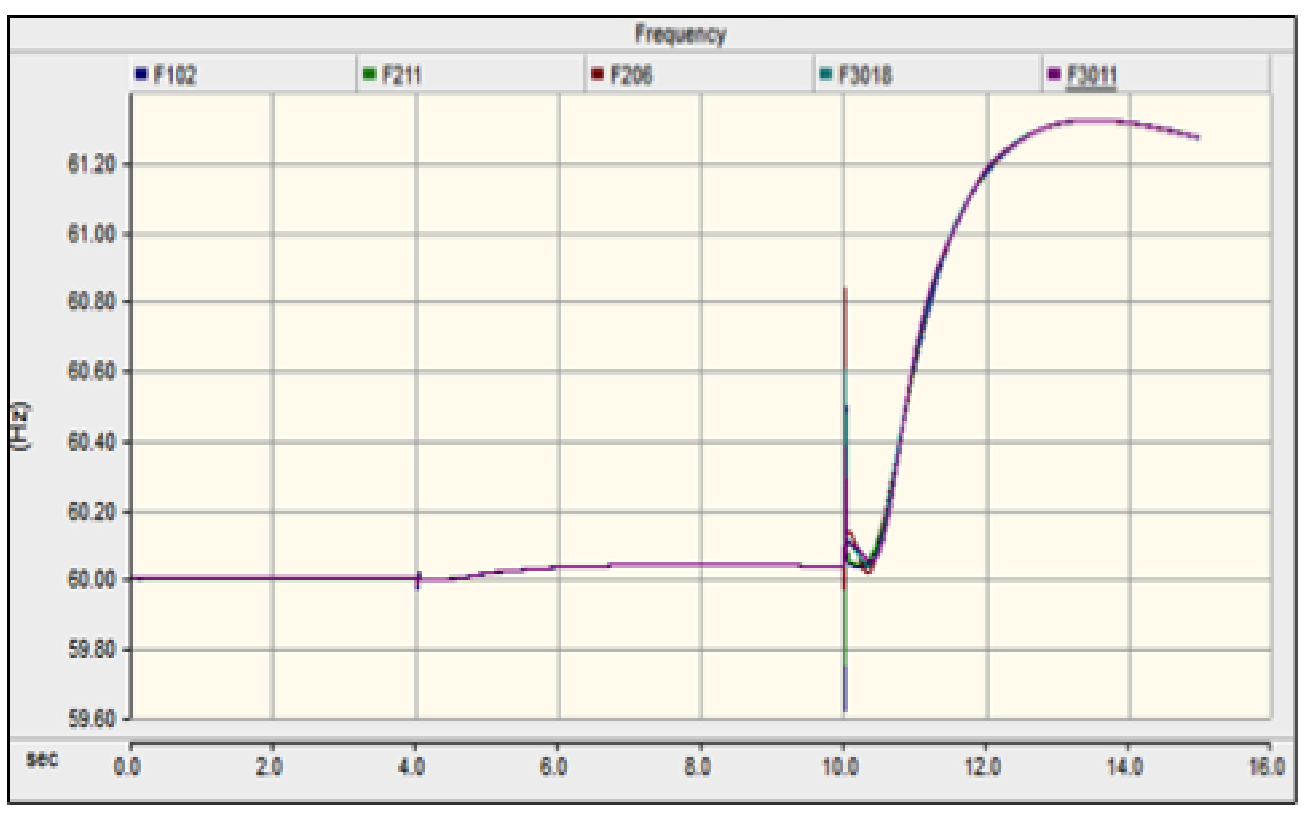

(B) Load Shedding Event

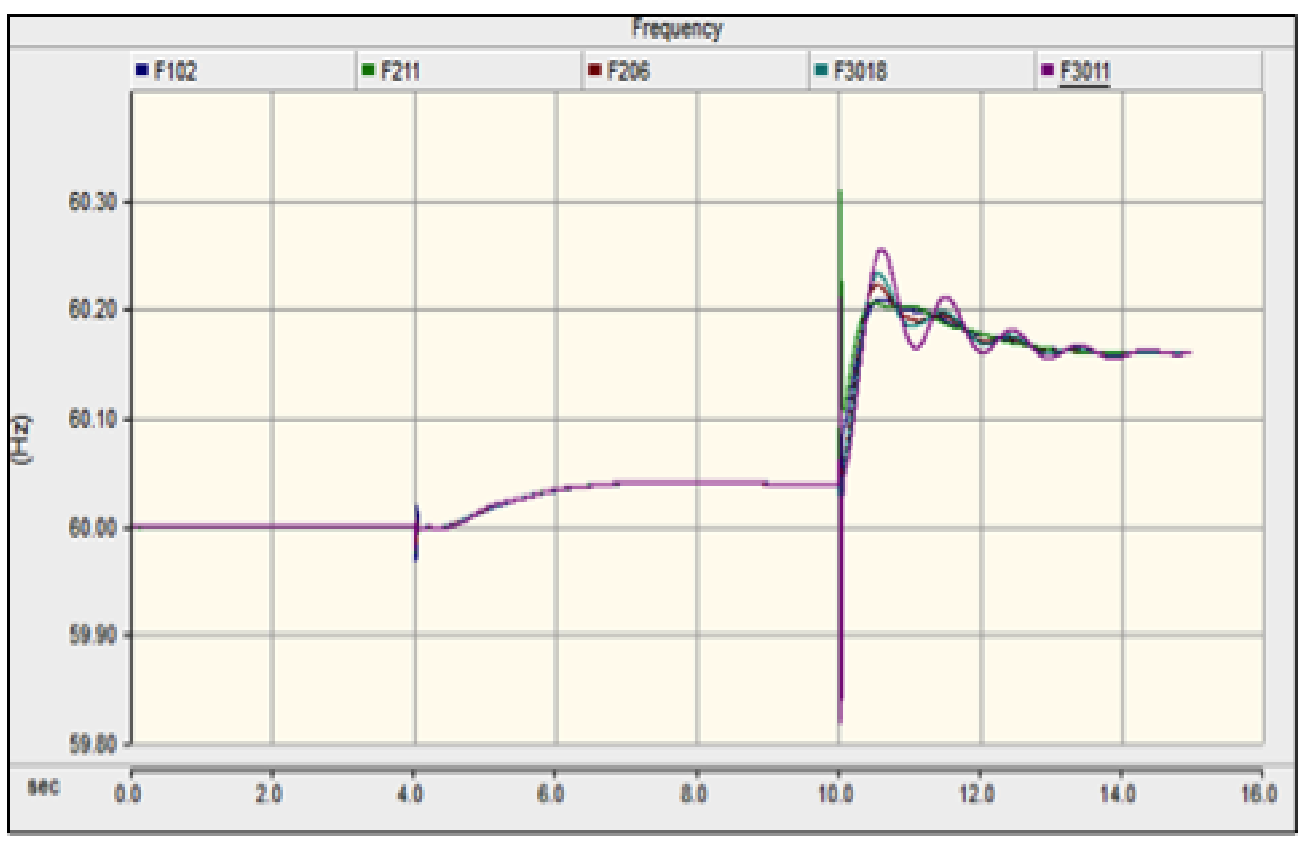

(C) Line Trip Event

Figure 14: Electromagnetic and electromechanical transients after events

In addition, it can be noticed that the transient as a result of different disturbance have different rising slopes. This discovery inspires students to explore more application possibilities based on electromechanical waves, an example of which is discussed in the following section.

## *D. Event Location Application Development*

More and more phasor measurement devices, such as PMU and Frequency Disturbance Recorder (FDR), are deployed in modern power systems. These devices can provide real-time phasor data for gird security applications. Simulation-based methods help to explore the possibility of new applications and also accelerate their development processes. This section uses the development of the event location application as an example. Event location is based on that electromechanical waves have relatively small error propagation velocity and regular propagation features. Here, the study of event location based on electromechanical waves is conducted on the NPCC model.

Supposing a generation trip event at Bus 133 occurs at 22 second in the NPCC system, an electromechanical wave will be formed and propagate to the entire system with a finite speed. The electromechanical transients measured at three locations are shown in Fig.15. The arrival time of electromechanical waves, which relates to the distance of measurement location to the event location, can be obtained from the transient waveforms. Using the least square method, the estimated and real locations of the generation trip event are shown in Fig.16.

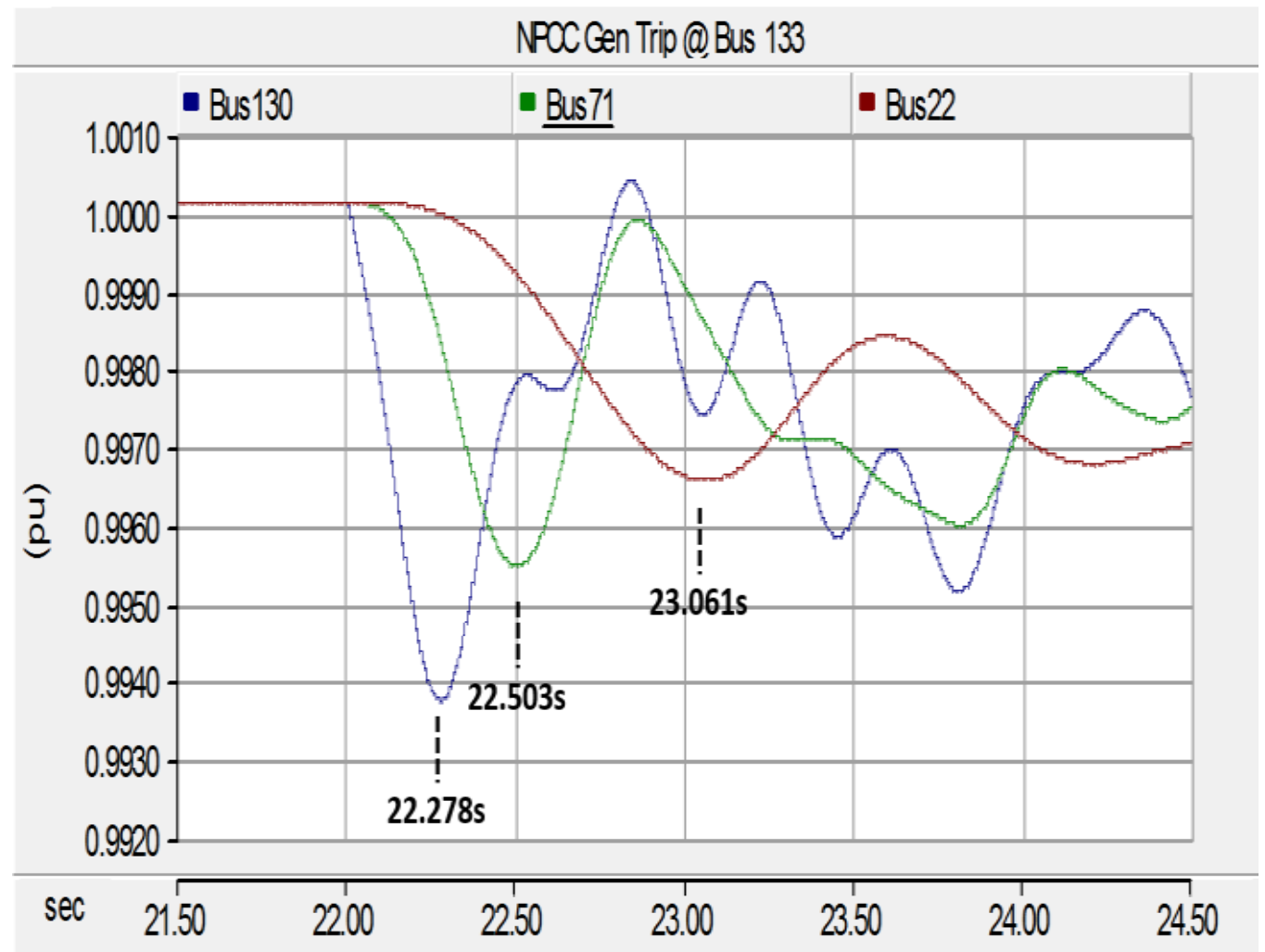

Figure 15: Electromechanical Transient (in P.U. Frequency)

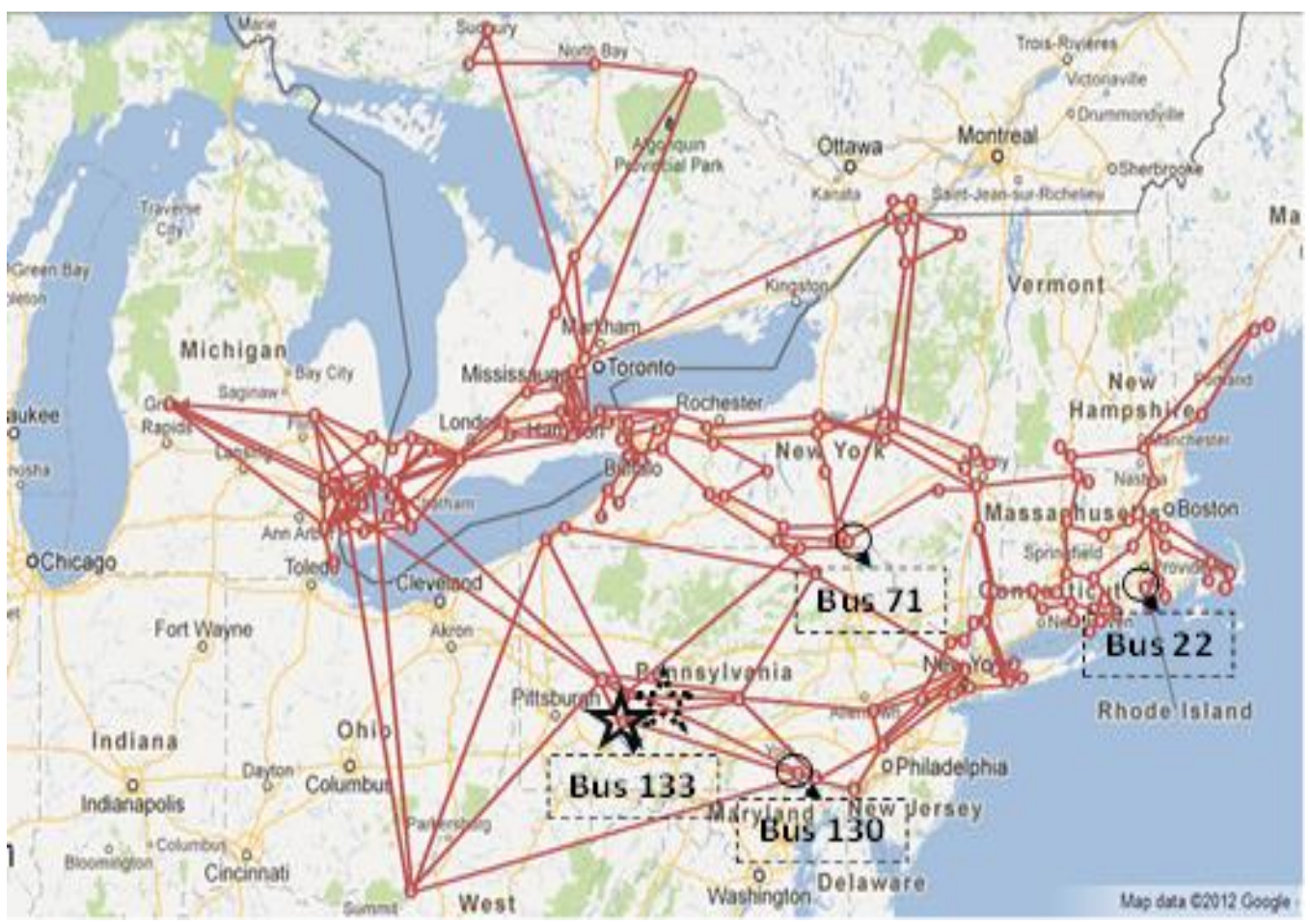

Figure 16: Event Location Based on Electromagnetic Waves.

The absolute error is 56 km, which is a relatively small error considering the scale of the system, showing that event location based on electromechanical waves is promising. This estimation error is due to that the wave propagation speed in the network is not necessarily the same in all directions. If the propagation speed of all directions can be calculated, the accuracy of event location based on electromechanical waves will greatly improve.

## V. Conclusion

Understanding the differences between electromagnetic and electromechanical waves in terms of focuses, modeling, and computation methods is an important step in power system transient and stability study. This paper introduced a simulation-based approach to study the two types of transient waves. This approach includes the development of three education models that can clearly show electromechanical and electromagnetic waves simultaneously in one single simulation case. Simulations with different settings on the developed models can visualize and verify the theoretical knowledge about the two types of transient waves. The propagation speed of both electromagnetic and electromechanical transient can be calculated theoretically and compared with simulations results based on the developed models. The models can also be used to investigate the changes of electromechanical propagation speed with machine inertia and transmission line reactance and the change of electromagnetic wave speed with the transmission line capacitance. The developed models can also help to learn features of typical electromechanical and electromagnetic waves in response to common disturbances such as generator trip, line trip, and load shedding. Using the propagation of the electromechanical waves, the process of developing an event location application is shown based on the developed models.